\documentclass[%
 reprint,
 superscriptaddress,
 amsmath,amssymb,
 aps,twocolumn
pra,floatfix,
pra,
]{revtex4-1}

\usepackage{graphicx}
\usepackage{dcolumn}
\usepackage{bm}
\usepackage{dsfont}
\usepackage{blindtext}

\newcommand\varpm{\mathbin{\vcenter{\hbox{%
  \oalign{\hfil$\scriptstyle+$\hfil\cr
          \noalign{\kern-.3ex}
          $\scriptscriptstyle({-})$\cr}%
}}}}

\begin{document}


\title{Distributing Memory Effects in an Open Two-Qubit System}

\author{Olli Siltanen}
\author{Tom Kuusela}
\author{Jyrki Piilo}
\affiliation{Turku Centre for Quantum Physics, Department of Physics and Astronomy, University of Turku, FI-20014 Turun yliopisto, Finland}
\affiliation{QTF Centre of Excellence, Department of Physics and Astronomy, University of Turku, FI-20014 Turun yliopisto, Finland}




\date{\today}

\begin{abstract}
We introduce a method to distribute memory effects among different subspaces of an open two-qubit system's state space. Within the linear optical framework, our system of interest is the polarization of two photons, while the environment consists of their frequency degrees of freedom. By exploiting initially correlated  frequency distributions and  initial frequency-dependent phase factors, we are able to control all the four decoherence functions for polarization -- corresponding to different open system subspace divisions --  almost independently of each other. Hence, our results demonstrate how, in a multipartite dephasing system, Markovian and non-Markovian dynamics can be arbitrarily divided between the subsystems, giving rise to a novel concept of memory partitions. We also discuss further implications the results have, e.g., on the dynamics of purity and entropy within the two-qubit system.
\end{abstract}

\maketitle



\section{\label{sec:intro}Introduction}
The studies of open quantum systems are important both for practical and fundamental reasons. The interaction between an open system and its environment leads to decoherence~\cite{petru} which needs to be accounted for when constructing practical quantum devices~\cite{suter}, e.g., for quantum computation. For fundamental aspects, there has recently been significant development in particular in understanding how non-Markovian memory effects influence the dynamics of open quantum systems~\cite{RevRivas,RevRMP,devega,LiRev,p1,p2} - and also how to engineer and exploit them for various tasks including quantum information framework~\cite{p2,telep,SDC,josza}.  
For example, non-Markovian dynamics has been demonstrated to enhance the performance of both superdense 
coding~\cite{SDC} and Deutsch-Jozsa algorithm~\cite{josza}.

Despite of the recent general progress on characterizing and quantifying non-Markovian effects, most of the systems studied in more detail within this open system framework often consist of exemplary single qubit systems. Whilst these works have illustrated important dynamical features, such as Markovian to non-Markovian transition~\cite{from_M_to_NM} and essentially arbitrary control of single qubit dephasing dynamics~\cite{dephasing_control}, extending the state space to multi-qubit systems allows to study a whole new set of problems
(see, e.g.,~\cite{tempo}).
Here, our interest lies on the problematics of how to distribute memory effects between the parties of a multi-party setting and whether this can be done in an arbitrary way - or do there exist fundamental limitations how this can be done?
According to a traditional view point, increasing the dimensionality of the open system should turn non-Markovian features to Markovian ones, when increasing the number of the degrees of freedom of the open system -- and at the same time decreasing those of the environment within the considered total system
~\cite{NM_to_M_1,NM_to_M_2,NM_to_M_3}.
However, we will show that also the opposite can happen -- and as a matter of fact -- it is possible to distribute memory effects arbitrarily between the different subspaces when considering two-qubit 
dephasing dynamics.   

Recent works have often used linear optical systems to study controlled open system dynamics and
their applications~\cite{from_M_to_NM,dephasing_control,Chiuri,Cialdi,fff,bnk,spectra,collision} -- and this is also our choice of physical platform for the current purpose.
Within this setting, the polarization degree of freedom of photons is the open system, while their frequency plays the role of an environment~\cite{from_M_to_NM}. The system and environment, in this context, are coupled in a birefringent medium such as quartz or calcite. The interaction due to this coupling manifests itself by  reducing the magnitude of  the off-diagonal elements of the system's density matrix while keeping the probability terms intact. Thereby, this corresponds to pure decoherence or dephasing. For a single photon, full control of the dephasing dynamics was recently achieved by manipulating frequency distribution of the photon and a given frequency dependent complex phase factor~\cite{dephasing_control}.  For a multi-partite setting, we show how these ingredients allow not only the control of the dephasing dynamics but also leads to
arbitrary control of the distribution of memory effects between different subsystems and subspaces of the open system.
In the current scheme, we also exploit initial correlations between the environments, which have been previously shown to lead to non-local memory effects~\cite{nonlocal_memory,photonic_real}.
 
 The paper is organized in the following way.
 Section~\ref{sec:SE} introduces the basics of dephasing for single- and two-qubit systems in linear optics.
 The core results of the paper are included in Section~\ref{sec:memory} which describes the subspace (or subparty) divisions, 
 shows explicitly how to distribute the memory effects, and discusses a number of other implications that our results have. 
 Section~\ref{sec:conclusion} concludes the paper.
\section{\label{sec:SE}Bipartite open system in linear optics}

The most general, initial and pure, \textit{one}-photon polarization-frequency state can be written as
\begin{equation}
\begin{split}
&|\Psi\rangle=\\
&a|H\rangle\int d\omega g(\omega)e^{i\theta_H(\omega)}|\omega\rangle+b|V\rangle\int d\omega g(\omega)e^{i\theta_V(\omega)}|\omega\rangle,\\
\end{split}
\label{1_photon}
\end{equation}
where $a$ and $b$ are the probability amplitudes for the photon to be in the polarization states $|H\rangle$ and $|V\rangle$, respectively, $H$ ($V$) stands for horizontal (vertical) polarization, $g(\omega)$ is the probability amplitude for the photon to be in the (angular) frequency state $|\omega\rangle$, and $e^{i\theta_{H(V)}(\omega)}$ is the complex phase factor corresponding to horizontal (vertical) polarization. Note that the phase doesn't need to be constant. 
Then, for \textit{two} photons, we can write
\begin{equation}
\begin{split}
&|\Psi\rangle=\\
&a|HH\rangle\int d\omega_1d\omega_2g(\omega_1,\omega_2)e^{i\theta_{1,H}(\omega_1)}e^{i\theta_{2,H}(\omega_2)}|\omega_1\rangle|\omega_2\rangle\\
&+b|HV\rangle\int d\omega_1d\omega_2g(\omega_1,\omega_2)e^{i\theta_{1,H}(\omega_1)}e^{i\theta_{2,V}(\omega_2)}|\omega_1\rangle|\omega_2\rangle\\
&+c|VH\rangle\int d\omega_1d\omega_2g(\omega_1,\omega_2)e^{i\theta_{1,V}(\omega_1)}e^{i\theta_{2,H}(\omega_2)}| \omega_1\rangle|\omega_2\rangle \\
&+d|VV\rangle\int d\omega_1d\omega_2g(\omega_1,\omega_2)e^{i\theta_{1,V}(\omega_1)}e^{i\theta_{2,V}(\omega_2)}|\omega_1\rangle|\omega_2\rangle.
\end{split}
\label{state}
\end{equation}
The parameters are to be interpreted as in the case of one photon, the lower indices referring to the photons 1 and 2, and the probability amplitudes satisfying $|a|^2+|b|^2+|c|^2+|d|^2=1$ and $\int d\omega_1d\omega_2|g(\omega_1,\omega_2)|^2=1$. For the sake of computational simplicity and experimental realizability, we have assumed that the phase factors are not correlated, i.e., $e^{i\theta_\lambda(\omega_1,\omega_2)}=e^{i\theta_{1,\lambda}(\omega_1)}e^{i\theta_{2,\lambda}(\omega_2)}$. To the best of our knowledge, non-local operations creating such correlations that $e^{i\theta_\lambda(\omega_1,\omega_2)}\neq e^{i\theta_{1,\lambda}(\omega_1)}e^{i\theta_{2,\lambda}(\omega_2)}$ haven't been reported even though it is possible to control the phase factors for single photons locally~\cite{dephasing_control}.

The system and environment are assumed to interact only locally. In practice, the two are coupled in birefringent media. As the local Hamiltonians are of the form
\begin{equation}
\mathcal{H}_j=\Big(n_{j,H}|H_j\rangle\langle H_j|+n_{j,V}|V_j\rangle\langle V_j|\Big)\int d\omega_j\omega_j|\omega_j\rangle\langle\omega_j|,
\label{Hamiltonian}
\end{equation}
where $n_{j,H(V)}$ is the refractive index of the $j$th photon corresponding to horizontal (vertical) polarization, the following local unitary dynamics apply:
\begin{equation}
U_j(t)|\lambda_j\rangle|\omega_j\rangle=e^{in_{j,\lambda}\omega_jt}|\lambda_j\rangle|\omega_j\rangle.
\label{local_unitary}
\end{equation}

The evolution of the bipartite open system is described by the dynamical map obtained by tracing out the environmental degrees of freedom
\begin{equation}
\begin{split}
\Phi_{12}(t)&(\varrho_{12}(0))=\\
&\text{tr}_E\Big[(U_1(t)\otimes U_2(t))|\Psi\rangle\langle\Psi|(U_1(t)^\dagger\otimes U_2(t)^\dagger)\Big].
\end{split}
\label{map}
\end{equation}
The corresponding density matrix of the open system at time $t$ is
\begin{equation}
\renewcommand{\arraystretch}{1.3}
\varrho_{12}(t)=
\begin{pmatrix}
|a|^2&ab^*\kappa_2(t)&ac^*\kappa_1(t)&ad^*\kappa_{12}(t)\\
a^*b\kappa^*_2(t)&|b|^2&bc^*\Lambda_{12}(t)&bd^*\kappa_1(t)\\
a^*c\kappa^*_1(t)&b^*c\Lambda^*_{12}(t)&|c|^2&cd^*\kappa_2(t)\\
a^*d\kappa^*_{12}(t)&b^*d\kappa^*_1(t)&c^*d\kappa^*_2(t)&|d|^2
\end{pmatrix},
\label{matrix}
\end{equation}
where
\begin{align}
\kappa_j(t)&=\int d\omega_1d\omega_2|g(\omega_1,\omega_2)|^2e^{i\theta_j(\omega_j)}e^{i\Delta n\omega_jt}\label{kappa_j},\\
\kappa_{12}(t)&=\int d\omega_1d\omega_2|g(\omega_1,\omega_2)|^2e^{i(\theta_1(\omega_1)+\theta_2(\omega_2))}e^{i\Delta n(\omega_1+\omega_2)t}\label{kappa_12},\\
\Lambda_{12}(t)&=\int d\omega_1d\omega_2|g(\omega_1,\omega_2)|^2e^{i(\theta_1(\omega_1)-\theta_2(\omega_2))}e^{i\Delta n(\omega_1-\omega_2)t}\label{lambda_12}
\end{align}
are the \textit{decoherence functions}. Above, we have denoted $\theta_j(\omega_j)=\theta_{j,H}(\omega_j)-\theta_{j,V}(\omega_j)$ and $\Delta n=n_{j,H}-n_{j,V}$, i.e., the birefringent media are assumed to be the same for both photons.
Note that initial polarization-frequency correlations are introduced when $\theta_j(\omega_j)$ is not constant.
Therefore, the magnitude of the decoherence functions at initial point of time can be less than $1$  and the domain of initial polarization states is restricted. However, one can use the rescaling $|\kappa_j(t)|/|\kappa_j(0)|$ -- and in similar manner for other decoherence functions -- to obtain the correspondence with completely positive dynamical map (see also the corresponding discussion in Ref.~\cite{dephasing_control}).

Accounting for initial correlations between the two photons' frequencies, described by $K\in[-1,1]$, we use the bivariate Gaussian
\begin{equation}
\begin{split}
|g(\omega_1,\omega_2)|^2&=\frac{1}{2\pi\sigma^2\sqrt{1-K^2}}e^{-\frac{(\omega_1-\mu)^2-2K(\omega_1-\mu)(\omega_2-\mu)+(\omega_2-\mu)^2}{2\sigma^2(1-K^2)}}\\
&=:G(K)
\end{split}
\label{Gauss}
\end{equation}
as the frequency distribution. $\mu$ and $\sigma^2$ are, respectively, the mean frequency and variance of both $\omega_1$ and $\omega_2$. The simple and experimentally realizable form of $G(K)$ -- also utilized, e.g., in~\cite{photonic_real,nonlocal_memory,SDC} -- eases the analysis of the frequency correlations.
$K=-1$ indicates perfect anticorrelation, i.e., the total frequency of the two photons being fixed: $\omega_1+\omega_2=2\mu$. Although perfect correlation is experimentally much more difficult to implement, we include it for the sake of theoretical generality. Perfectly correlated ($K=1$) frequencies satisfy $\omega_1=\omega_2$. Both of the extreme cases have a useful property: $K=\varpm1$ results in the magnitude of the decoherence function $\Lambda_{12}(t)$ $(\kappa_{12}(t))$ being constant. Having no initial correlations ($K=0$) factorizes the dynamics. That is, $\kappa_{12}(t)=\kappa_1(t)\kappa_2(t)$, $\Lambda_{12}(t)=\kappa_1(t)\kappa_2^*(t)$, and $\Phi_{12}(t)=\Phi_1(t)\otimes\Phi_2(t)$. Furthermore, if the phase distributions $\theta_j(\omega_j)$ were constant, the coherence terms of Eq.~\eqref{matrix} would experience
a Gaussian
decay, obeying the local decoherence functions
\begin{equation}
\kappa_j(\tau)=e^{-\frac{1}{2}\tau^2+i(\eta\tau+\theta_j)},
\label{dephasing}
\end{equation}
where we have denoted $\tau=\sigma\Delta nt$ and $\eta=\mu/\sigma$. To go beyond the dynamics predicted by Eq.~\eqref{dephasing}, we modify $|g(\omega_1,\omega_2)|^2$ and $\theta_j(\omega_j)$ to engineer the dephasing dynamics and ultimately to control the distribution of memory effects.

The frequency distribution of the photon pairs depends on the spectrum of the pump laser and focusing conditions on the non-linear crystal, and the cases of $K=-1$ and $K=0$ are easily realized. According to numerical simulations positive correlation ($K\approx1$) might be possible with a broadband femtosecond laser impinging a very long crystal~\cite{K_approx_one}. Furthermore, interference filters and Fabry-Perot cavities can be used to manipulate the spectrum of individual photons~\cite{from_M_to_NM}. Full control of the dephasing dynamics is also already possible. It has been demonstrated that excellent control of both the local frequency distribution and differential complex phase of both polarization components is achieved by spatial light modulators~\cite{dephasing_control}. And finally, the total interaction time is adjusted by changing the thickness of the quartz plates which couple the polarization with frequency.

\begin{figure}[t]
\centering
\includegraphics[width=.6\linewidth]{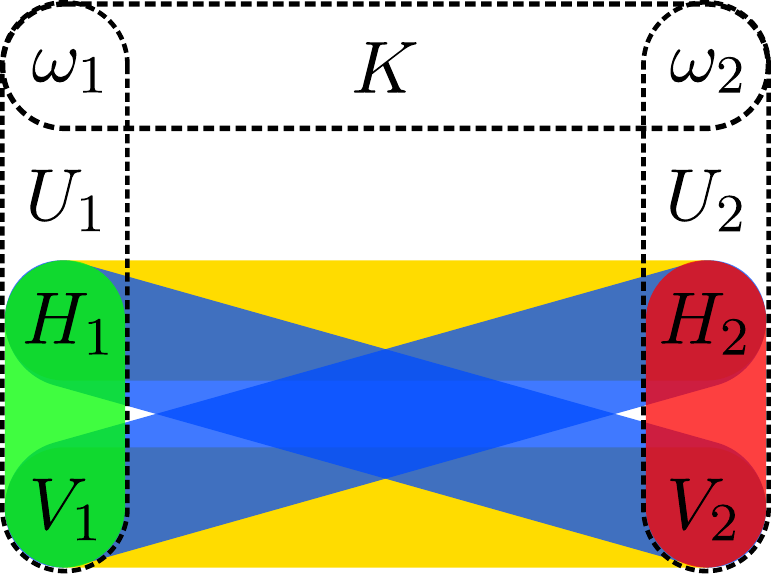}
\caption{(Color online) The environment degrees of freedom $\omega_j$ coupled with the system degrees of freedom $H_j$ and $V_j$ by the local unitaries $U_j$, where $j=1,2$. $K$ describes the initial correlations between the two environments. Different subspaces are illustrated by the ovals of different color. Green = $\mathbf{S}_1$, red = $\mathbf{S}_2$, yellow = $\mathbf{S}_\Phi$, blue = $\mathbf{S}_\Psi$.}
\label{partition_scheme}
\end{figure}


\section{\label{sec:memory}Distributing memory effects among different subspaces}


\subsection{\label{subsec:mp_definition}Memory effects in subspaces}

The decoherence functions \eqref{kappa_j}--\eqref{lambda_12} describe the dephasing dynamics of states in the subspaces
\begin{align}
&\mathbf{S}_j=\operatorname{span}\big(\{|H_j\rangle,|V_j\rangle\}\big),\\
&\mathbf{S}_\Phi=\operatorname{span}\big(\{|\Phi^+\rangle,|\Phi^-\rangle\}\big),\\
&\mathbf{S}_\Psi=\operatorname{span}\big(\{|\Psi^+\rangle,|\Psi^-\rangle\}\big),
\end{align}
respectively, where $j=1,2$, $|\Phi^\pm\rangle=\frac{1}{\sqrt{2}}(|HH\rangle\pm|VV\rangle)$, and $|\Psi^\pm\rangle=\frac{1}{\sqrt{2}}(|HV\rangle\pm|VH\rangle)$. These subspaces, as well as the open system dynamics, are illustrated in Fig.~\ref{partition_scheme}.

Since we are  dealing with pure decoherence, whether or not memory effects can be associated with a certain subspace, is determined by the monotonicity of the corresponding decoherence function's magnitude. That is, there are memory effects in the subspace $\mathbf{S}_j$ if $|\kappa_j(t)|$ behaves in a non-monotonic manner, and similarly for the subspace $\mathbf{S}_\Phi$ $\left(\mathbf{S}_\Psi\right)$ and the decoherence function $|\kappa_{12}(t)|$ $\left(|\Lambda_{12}(t)|\right)$.

\begin{table*}[t]
\caption{Frequency distributions, phase distributions and the resulting memory partitions.}
\begin{ruledtabular}
\begin{tabular}{ccccc}
Subfigure of Fig.~\ref{16_plots} & Frequency dist. & Phase dist. 1 & Phase dist. 2 & $\langle\mathbf{S}_1,\mathbf{S}_2,\mathbf{S}_\Phi,\mathbf{S}_\Psi\rangle$\\
\hline
a & $G(0)$ & constant & constant & $\langle0,0,0,0\rangle$\\
b & $G(0)$ & $z(\omega_1,5)$ & constant & $\langle1,0,0,0\rangle$\\
c & $G(0)$ & constant & $z(\omega_2,5)$ & $\langle0,1,0,0\rangle$\\
d & $[G(1)+G(-1)]/2$ & $p(\omega_1,3)$ & $p(\omega_2,3)$ & $\langle0,0,1,0\rangle$\\
e & $[G(1)+G(-1)]/2$ & $p(\omega_1,3)$ & $p(\omega_2,-3)$ & $\langle0,0,0,1\rangle$\\
f & $[G(1)+G(-1)]/2$ & $z(\omega_1,5)$ & $z(\omega_2,-5)$ & $\langle1,1,0,0\rangle$\\
g & $G(1)$ & $z(\omega_1,5)$ & constant & $\langle1,0,1,0\rangle$\\
h & $G(-1)$ & $z(\omega_1,5)$ & constant & $\langle1,0,0,1\rangle$\\
i & $G(1)$ & constant & $z(\omega_2,5)$ & $\langle0,1,1,0\rangle$\\
j & $G(-1)$ & constant & $z(\omega_2,5)$ & $\langle0,1,0,1\rangle$\\
k & $[G(1)+G(-1)]/2$ & $p(\omega_1,3)$ & constant & $\langle0,0,1,1\rangle$\\
l & $G(1)$ & $z(\omega_1,5)$ & $z(\omega_2,-15)$ & $\langle1,1,1,0\rangle$\\
m & $G(-1)$ & $z(\omega_1,5)$ & $z(\omega_2,-15)$ & $\langle1,1,0,1\rangle$\\
n & $G(0)$ & $z(\omega_1,5)$ & $p(\omega_2,3)$ & $\langle1,0,1,1\rangle$\\
o & $G(0)$ & $p(\omega_1,3)$ & $z(\omega_2,5)$ & $\langle0,1,1,1\rangle$\\
p & $G(0)$ & $z(\omega_1,5)$ & $z(\omega_2,5)$ & $\langle1,1,1,1\rangle$\\
\end{tabular}
\end{ruledtabular}
\label{specs}
\end{table*}

By controlling the decoherence functions and their monotonicity, we can control what combination of the subspaces is exposed to memory effects. Hence, we can talk about distributing memory effects. Since the monotonicity of a decoherence function is independent of the monotonicity of the other decoherence functions (at least in the case of two qubits, as we will show), there are $\sum_{k=0}^4\binom{4}{k}=16$ ways to distribute memory effects among the quadruplet $\langle\mathbf{S}_1,\mathbf{S}_2,\mathbf{S}_\Phi,\mathbf{S}_\Psi\rangle$, one of which corresponds to the trivial case of not having memory effects at all. We call these quadruplets memory partitions and write 1 in the place of those subspaces that experience the above described memory effects. The rest of the subspaces are indicated by 0 (see Table \ref{specs}). For example, the notation $\langle1,0,0,0\rangle$ is reserved for the memory partition where only $|\kappa_1(t)|$ behaves non-monotonically, and only the local states 
\begin{equation}
\renewcommand{\arraystretch}{1.3}
\varrho_{1}(t)=
\begin{pmatrix}
|a|^2+|b|^2&(ac^*+bd^*)\kappa_1(t)\\
(a^*c+b^*d)\kappa^*_1(t)&|c|^2+|d|^2
\end{pmatrix},
\label{local_matrix}
\end{equation}
experience memory effects. Moreover, in the case of $\langle1,1,0,0\rangle$ ($\langle0,0,1,1\rangle$), we say that the local dephasing dynamics is non-Markovian (Markovian), and the non-local dynamics is Markovian (non-Markovian).


\subsection{\label{subsec:decoherence}Controlling the decoherence functions}

Combining the results for the phase control~\cite{dephasing_control} (now for two photons instead of one) and frequency correlation control~\cite{nonlocal_memory} allows us to create different memory partitions. That is, we choose the phase distributions $\theta_j(\omega_j)$ and the correlation coefficient $K$ appropriately. As we aim to solve the integrals \eqref{kappa_j}--\eqref{lambda_12} analytically, we find, by trial and error, that as simple functions as the ``zigzag" function
\begin{equation}
z(\omega_j,\alpha_j)=\arcsin\Big[\sin\Big(\alpha_j\frac{\omega_j-\mu}{\sigma}\Big)\Big]
\label{zigzag}
\end{equation}
and the parabola
\begin{equation}
p(\omega_j,\beta_j)=\beta_j\Big(\frac{\omega_j-\mu}{\sigma}\Big)^2
\label{parabola}
\end{equation}
not only do this but are enough for us to obtain all the 15 non-trivial memory partitions, when used as the phase distributions and accompanied by a suitable Gaussian frequency distribution.

The zigzag phase, when substituted into the decoherence functions \eqref{kappa_j}--\eqref{lambda_12}, yields multiple local (and depending on the value of the correlation coefficient, also non-local) ``recoherence" peaks. That is, the magnitudes of the selected decoherence functions increase after the initial drops, and then decrease again.  The distance between these peaks is controlled by $\alpha_j$, i.e., the angular frequency of the zigzag wave [see Eqs. \eqref{zigzag_kappa_j}--\eqref{zigzag_nonlocal_constant_deco} in Appendix A for details].

\begin{figure*}[t]
\begin{minipage}{\textwidth}
\includegraphics[width=\textwidth]{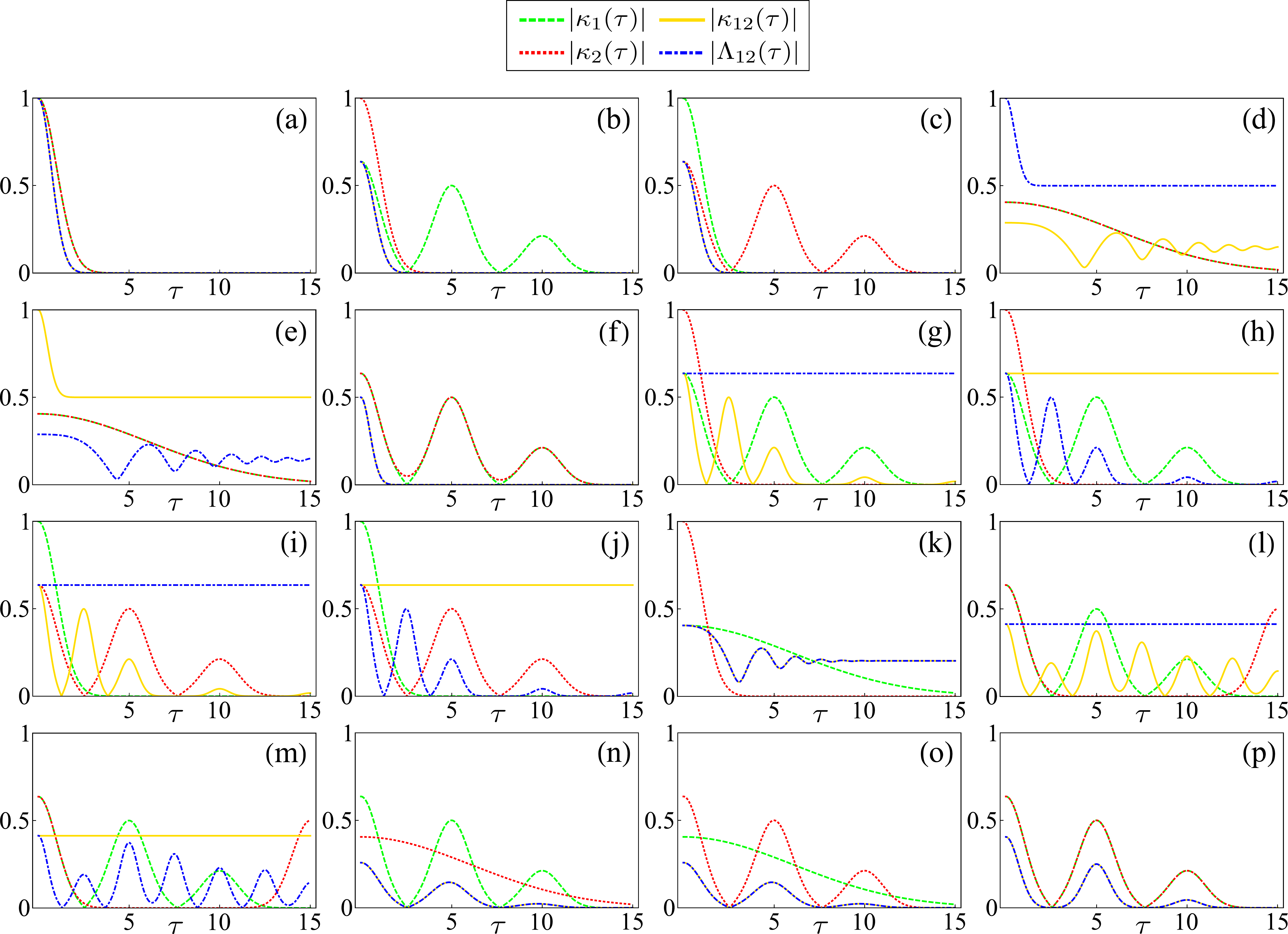}
\caption{(Color online) Magnitudes of the decoherence functions creating all 16 memory partitions as functions of the scaled, unitless time $\tau=\sigma\Delta nt$. See Table~\ref{specs} for used frequency and phase distributions, and corresponding memory partitions, for panels (a)--(p).}
\label{16_plots}
\end{minipage}
\end{figure*}


The parabola phase transforms the absolute values of the decoherence functions into Gaussian functions [see Eqs. \eqref{abs_parabola_kappa_j}--\eqref{parabola_lambda_12} in Appendix B]. Using the parabola phase, one achieves no local recoherence but is able to control the height and width of the Gaussians by the parameter $\beta_j$, i.e., the inverse of the parabola's focal width. Moreover, we gain non-local memory effects by implementing the frequency distribution $\frac{1}{2}[G(1)+G(-1)]$ [see Eqs. \eqref{abs_parabola_kappa_12}, \eqref{abs_parabola_lambda_12}  in Appendix B], motivated by simultaneous need for both extremes; The frequencies of the photons 1 and 2, described by this distribution, are equally probable to be perfectly correlated or anticorrelated.

If neither the zigzag nor the parabola phase is required in obtaining a certain memory partition, we employ an arbitrary constant phase factor. The choices of frequency distributions, phase distributions, and the resulting memory partitions are depicted in Table \ref{specs}, while the corresponding magnitudes of the decoherence functions are plotted in Fig.~\ref{16_plots}. Here, we do not address the question, which intervals of $\alpha_j$ and $\beta_j$ produce memory effects. Instead,
based on a numerical analysis,
we employ the example values $\alpha_j\in\{-15,\pm5\}$ and $\beta_j\in\{\pm3\}$.
It is also worth mentioning, that the decoherence functions alone do not determine the dephasing dynamics of a given system in general, but one also needs to consider the initial probability amplitudes $a$, $b$, $c$, and $d$ of the bipartite polarization state.


\subsection{\label{subsec:deriving}Deriving the memory partitions}

The trivial partition $\langle0,0,0,0\rangle$ [Fig.~\ref{16_plots} (a)] is obtained by Eq.~\eqref{dephasing} and, in the case of the single-peaked Gaussian \eqref{Gauss}, any value of $K$. Here, we have chosen $K=0$.

To equip either one of the local subspaces $\mathbf{S}_j$ with memory effects [Fig.~\ref{16_plots} (b, c)], we make the substitution $\theta_j(\omega_j)=z(\omega_j,5)$, while fixing the other phase constant and $K=0$. Clearly, both of the non-local decoherence functions are non-monotonic too, being products of a monotonic and a non-monotonic function. However, because we can force the revivals of $|\kappa_{12}(t)|$ and $|\Lambda_{12}(t)|$ arbitrarily close to 0
by increasing $\alpha_j$, we omit them. For example, when $\alpha_j=5(10)$, the height of the first recoherence peak is of the order of $10^{-4}(10^{-12})$.

To equip either one of the non-local subspaces with memory effects, we choose the parabola phases and either $\beta_1=\beta_2$ [$\langle0,0,1,0\rangle$, Fig.~\ref{16_plots} (d)] or $\beta_1=-\beta_2$ [$\langle0,0,0,1\rangle$, Fig.~\ref{16_plots} (e)]. In both cases, we use $\frac{1}{2}[G(1)+G(-1)]$ as the frequency distribution. The most striking memory partition $\langle0,0,1,1\rangle$ [Fig.~\ref{16_plots} (k)] is obtained
similarly. Here, $|\kappa_{12}(t)|=|\Lambda_{12}(t)|$, when $\beta_2=0$ (or $\theta_2(\omega_2)=$ constant). 
This demonstrates that each one-qubit subsystem behaves individually in a Markovian manner, whilst the combined two-qubit system displays non-Markovian memory effects.
All the three memory partitions described above are applicable in protecting the so-called X-states,
the states containing non-zero density matrix elements only on their main diagonals and anti-diagonals~\cite{x_state}, from dephasing.
The memory partitions ensure that the coherence terms of such states always stay well above zero, monotonic or not.

The memory partion $\langle1,1,0,0\rangle$ [Fig.~\ref{16_plots} (f)] is obtained by applying $\frac{1}{2}[G(1)+G(-1)]$ and zigzag phases of opposite sign $\alpha_j$s. Thus, we have locally non-Markovian but non-locally Markovian dynamics. From a technical point of view, the behavior of the decoherence functions in this case is not that difficult to understand; $z(\omega_1,\alpha_1)$ and $z(\omega_2,-\alpha_1)$, in $|\kappa_{12}(t)|$ ($|\Lambda_{12}(t)|$), cancel each other in the presence of perfect anticorrelation (correlation). However, having local memory effects in each of the subsystems that do not manifest themselves at all non-locally in the combined system, is somewhat counterintuitive.

The memory partitions $\langle i,i\oplus1,j,j\oplus1\rangle$ [$i,j=0,1$, Fig.~\ref{16_plots} (g--j)] are obtained by simply choosing the zigzag phase, when the corresponding decoherence function needs to behave non-monotonically, and $K=1$ ($-1$), when $|\kappa_{12}(t)|$ ($|\Lambda_{12}(t)|$) needs to behave non-monotonically.

Perfect correlation and anticorrelation are used as ``switches" when producing the memory partitions $\langle1,1,1,0\rangle$ and $\langle1,1,0,1\rangle$ [Fig.~\ref{16_plots} (l, m)] also. $K=1$ forces $|\Lambda_{12}(t)|$ monotonic (constant), and $K=-1$ forces $|\kappa_{12}(t)|$ monotonic (constant). Using zigzag phases with different values of $\alpha_j$ results in the remaining non-local decoherence function being initially decreasing and non-monotonic. Perhaps surprisingly, $\alpha_1=\alpha_2$ does not yield non-local memory effects.

The partitions $\langle i,i\oplus1,1,1\rangle$ [$i=0,1$, Fig.~\ref{16_plots} (n, o)] are obtained by fixing $K=0$, i.e., $|\kappa_{12}(t)|=|\Lambda_{12}(t)|=|\kappa_1(t)\kappa_2(t)|$. We use the zigzag phase in the usual manner: For $\mathbf{S}_j$ to experience memory effects, $\theta_j(\omega_j)=z(\omega_j,5)$. As for the other phase, we choose the parabola, $\theta_k(\omega_k)=p(\omega_k,3)$, $k\neq j$. Hence, the revivals of the non-local decoherence functions become much more distinct than with $\langle i,i\oplus1,0,0\rangle$ and cannot be omitted anymore. The fully non-Markovian partition $\langle1,1,1,1\rangle$ [Fig.~\ref{16_plots} (p)] is achieved by changing the parabola phase into a zigzag phase such that $\alpha_1=\alpha_2$.


\subsection{\label{subsec:implications}Properties of selected memory partitions}

We will now proceed to investigate some of the features of the more prominent memory partitions. Here, the partitions $\langle1,1,0,0\rangle$ and $\langle0,0,1,1\rangle$, when $a=b=c=d=\frac{1}{2}$, are particularly interesting, because memory effects have been equally distributed within the local and non-local parts of the system, respectively. Hence, we will focus on these situations.

\begin{figure}[t]
\centering
\includegraphics[width=.95\linewidth]{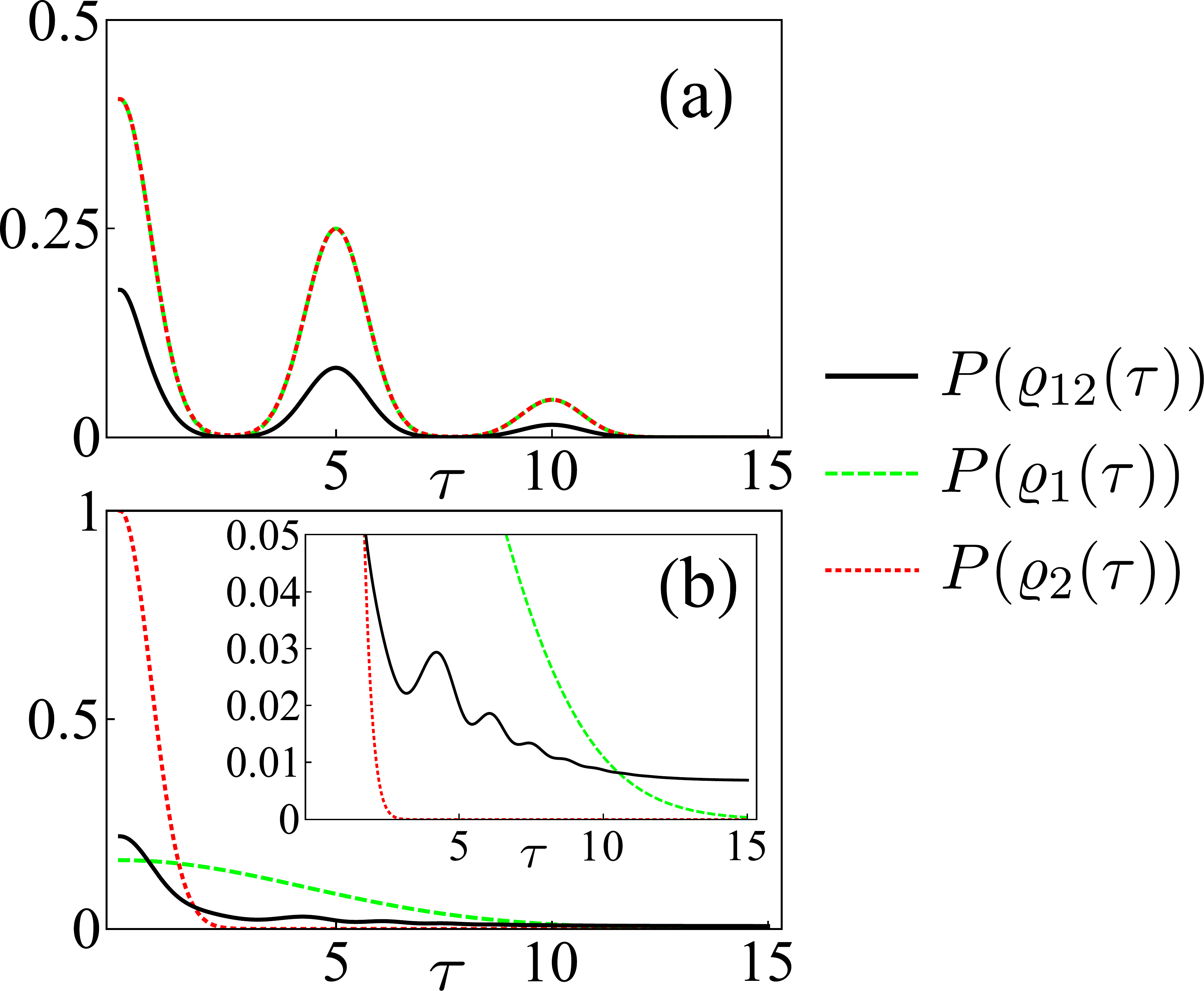}
\caption{(Color online) Normalized purities of the total state $\varrho_{12}(\tau)$ and its reduced states $\varrho_j(\tau)$, when $a=b=c=d=\frac{1}{2}$, in the case of (a) $\langle1,1,0,0\rangle$, and (b) $\langle0,0,1,1\rangle$. The inset shows a magnification of the same purities.}
\label{purities}
\end{figure}

Purity~\cite{teiko} describes the ``quantumness" of a given quantum state. Purities $P$ of the bipartite open system and its subsystems, normalized by $\frac{D}{D-1}(P-\frac{1}{D})$, where $D$ is the dimension of the system, are plotted in Fig.~\ref{purities}. The local decoherence functions clearly dictate the behavior of the local purities, but they also contribute to the total purity, as one can see especially from Fig.~\ref{purities} (a). Contrary to $\langle1,1,0,0\rangle$, with $\langle0,0,1,1\rangle$, the contribution of the non-local decoherence functions to the total purity becomes essential [Fig.~\ref{purities} (b)]. Unlike the local purities, the purity of the total state undergoes revivals and has a positive limit, resulting in a total system purer than its parts. In general, this is the case with decoherence-free subspaces as well~\cite{dfs1,dfs2}, which can be obtained by applying either $G(1)$ or $G(-1)$. However, depending on the phase distributions and the probability amplitudes, mixing the two may sometimes be more beneficial. That is, applying more noise can actually result in richer dynamics and, in some scenarios, purer states.

\begin{figure}[t]
\centering
\includegraphics[width=.95\linewidth]{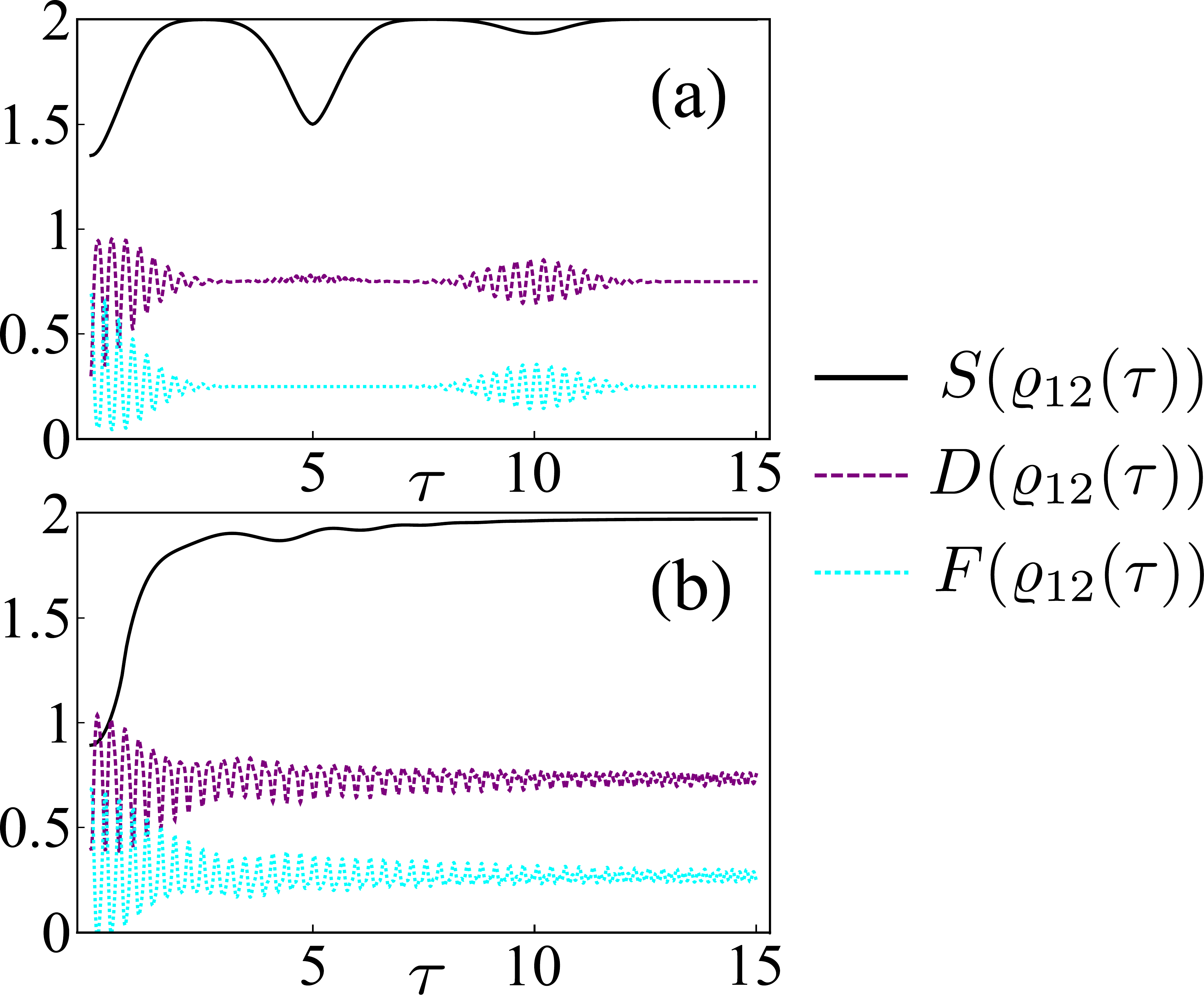}
\caption{(Color online) Entropy $S$, trace distance $D$ and fidelity $F$ of the total state $\varrho_{12}(\tau)$, the latter two with respect to the state $|++\rangle$, when $a=b=c=d=\frac{1}{2}$, in the case of (a) $\langle1,1,0,0\rangle$, and (b) $\langle0,0,1,1\rangle$. We have fixed $\theta_2=0$ and $\eta=20$.}
\label{S_D_F}
\end{figure}


As the coherence terms of the open bipartite system vanish (or at least decrease, on average), one might be tempted to ask what this implies for the total system. Now, since the coherences are associated with quantum correlations~\cite{quantum_correlation}, it is reasonable to address this question from the point of view of correlation flow. In the context of open quantum systems, increasing von Neumann entropy of the open system (plotted in Fig.~\ref{S_D_F}) adds to the total correlations between the system and environment~\cite{entropy}. Because the purity of a closed system, which the open system and its environment together form, is constant,  the coherences \textit{within the system} become correlations \textit{between the system and environment}, and 
vice versa. One can also see this from Figs. \ref{purities} and \ref{S_D_F}.

To see how the decohering open systems behave with respect to the corresponding closed systems, described by the pure state $|++\rangle=\frac{1}{2}(|HH\rangle+|HV\rangle+|VH\rangle+|VV\rangle)$, we have plotted their trace distances~\cite{chuang} and fidelities~\cite{fidelity} in Fig.~\ref{S_D_F} as well. The oscillation frequency of the trace distance and fidelity is determined by the mean frequency $\mu$ and the constant parts of the phase distributions. The oscillation amplitude, on the other hand, behaves not as straightforwardly but seems to be proportional to the purity. We now make two interesting observations. First, although correlations flow back to the system, that doesn't necessarily mean the bipartite open system gets any closer to its closed counterpart [Fig.~\ref{S_D_F} (a)]. Instead, it may get closer to some other pure state. In this particular case that state, around $\tau=5$, is $|+-\rangle=\frac{1}{2}(|HH\rangle-|HV\rangle+|VH\rangle-|VV\rangle)$. Secondly, we can control the limit of the oscillation amplitude and thus prevent the oscillations from ever coming to an end [Fig.~\ref{S_D_F} (b)].

\section{\label{sec:conclusion} Conclusions}
We have shown how to select and target subspaces of a bipartite open system to experience memory effects in a linear optical scheme. Indeed, the results demonstrate that by engineering the dephasing dynamics, any chosen combination of subspaces can undergo non-Markovian dynamics - including the case where each party individually follows Markovian evolution whilst the combined bipartite open system displays non-Markovian features. This leads to the concept of memory partitions in the dynamics of multi-partite open quantum systems.
The results also imply that one could quantify how common non-Markovian features in a multi-partite open systems are by calculating the fraction of subspaces displaying memory effects.

Extending from the two-partite case considered here to a generic $N$-qubit system, it remains to be shown  
whether a generic strategy of distributing memory effects for an arbitrary number of qubits exists. 
It is also worth noting that for $N$ qubits there are
$M:=\sum_{k=1}^N\binom{N}{k}2^{k-1}$ different decoherence functions and thus $2^M$ memory partitions. The number of decoherence functions is larger than $2^N=d_s$, the size of the open system's Hilbert space, when $d_s>4$. This may open up new possibilities to convey information when exploiting different memory partitions and choosing whether they carry memory effects or not. One could, e.g., encode information to the dynamics of the $N$-qubit system's coherence terms. The corresponding decoding protocol then would require at least two predetermined measurement points, were it not possible to decode a snapshot of the system's evolution.
We also expect that the techniques introduced here can be useful in various other tasks including, e.g., the partial protection of two-qubit X-states by introducing and engineering the memory effects in the corresponding two subspaces.
In general, engineering the open system dynamics is an important feature for several applications of quantum mechanics -- and our results open the path to manipulate the dynamics within each subspace individually when dealing with multi-partite open systems.



\section*{\label{sec:acknow}Acknowledgements}

O.S. acknowledges the useful discussions with Henri Lyyra. This work was financially supported by a grant for postgraduate studies from the Magnus Ehrnrooth Foundation and the Academy of Finland via the Centre of Excellence program (Project no. 312058).



\onecolumngrid

\subsection*{\label{sec:appendix_a}Appendix A: Solving the decoherence functions using the zigzag phase distribution}

\setcounter{equation}{0}
\renewcommand{\theequation}{A\arabic{equation}}

To solve Eqs. \eqref{kappa_j}--\eqref{lambda_12} using the zigzag phase \eqref{zigzag}, we exploit the following identities:

\begin{equation}
e^{if(x)}=\cos(f(x))+i\sin(f(x)),
\label{Euler}
\end{equation}
\begin{equation}
\begin{cases}
\cos(\arcsin(\sin(Ax)))=|\cos(Ax)|\\
\sin(\arcsin(\sin(Ax)))=\sin(Ax)
\end{cases},
\end{equation}
\begin{equation}
|\cos(Ax)|=\frac{2}{\pi}+\frac{4}{\pi}\sum_{n=1}^\infty\frac{(-1)^n}{1-4n^2}\cos(2nAx),
\label{Fourier}
\end{equation}
\begin{equation}
\int dx\frac{1}{\sqrt{2\pi}}e^{-\frac{1}{2}x^2}\cos(Ax)\cos(Bx)\cos(Cx)
=\frac{1}{4}e^{-\frac{1}{2}(A+B+C)^2}\Big(1+e^{2(A+B)C}+e^{2(B+C)A}+e^{2(C+A)B}\Big),
\label{integral_cos_cos_cos}
\end{equation}
\begin{equation}
\int dx\frac{1}{\sqrt{2\pi}}e^{-\frac{1}{2}x^2}\cos(Ax)\sin(Bx)\sin(Cx)
=\frac{1}{4}e^{-\frac{1}{2}(A+B+C)^2}\Big(-1+e^{2(A+B)C}-e^{2(B+C)A}+e^{2(C+A)B}\Big).
\label{integral_cos_sin_sin}
\end{equation}

In the following, as well as in Appendix B, we make the change of variable(s), $x_j=\frac{\omega_j-\mu}{\sigma}$. The initial correlations between the two environments cannot be detected locally. Hence, the local decoherence functions become

\begin{equation}
\begin{split}
\kappa_j(\tau)&=e^{i\eta\tau}\int dx_j\frac{1}{\sqrt{2\pi}}e^{-\frac{1}{2}x_j^2}e^{i\arcsin(\sin(\alpha_jx_j))}e^{i\tau x_j}\\
&=e^{i\eta\tau}\int dx_j\frac{1}{\sqrt{2\pi}}e^{-\frac{1}{2}x_j^2}(|\cos(\alpha_jx_j)|+i\sin(\alpha_jx_j))(\cos(\tau x_j)+i\sin(\tau x_j))\\
&=e^{i\eta\tau}\int dx_j\frac{1}{\sqrt{2\pi}}e^{-\frac{1}{2}x_j^2}(|\cos(\alpha_jx_j)|\cos(\tau x_j)-\sin(\alpha_jx_j)\sin(\tau x_j))\\
&=e^{i\eta\tau}\int dx_j\frac{1}{\sqrt{2\pi}}e^{-\frac{1}{2}x_j^2}\Big[\Big(\frac{2}{\pi}+\frac{4}{\pi}\sum_{n=1}^\infty\frac{(-1)^n}{1-4n^2}\cos(2n\alpha_jx_j)\Big)\cos(\tau x_j)-\sin(\alpha_jx_j)\sin(\tau x_j)\Big]\\
&=e^{i\eta\tau}\Big[\frac{2}{\pi}e^{-\frac{1}{2}\tau^2}+\frac{2}{\pi}\sum_{n=1}^\infty\frac{(-1)^n}{1-4n^2}\Big(e^{-\frac{1}{2}(\tau-2n\alpha_j)^2}+e^{-\frac{1}{2}(\tau+2n\alpha_j)^2}\Big)-\frac{1}{2}\Big(e^{-\frac{1}{2}(\tau-\alpha_j)^2}-e^{-\frac{1}{2}(\tau+\alpha_j)^2}\Big)\Big].
\end{split}
\label{zigzag_kappa_j}
\end{equation}

\noindent We solve the non-local decoherence functions using only the extreme values of $K$. Hence, we can use the normal distribution of a single variable ($x_1$) as the frequency distribution, and substitute $x_2=Kx_1$ into the phase and time dependent part of the second photon. Skipping the lengthy details, the resulting decoherence functions are

\begin{equation}
\begin{split}
&\kappa_{12}(\tau)\big\vert_{K=1}=e^{i2\eta\tau}\Lambda_{12}(\tau)\big\vert_{K=-1}\\
&\hspace{48.5pt}=e^{-2\tau^2+i2\eta\tau}\Big\{\frac{4}{\pi^2}-\frac{1}{\pi}\Big(e^{-\frac{1}{2}\alpha_1(\alpha_1-4\tau)}+e^{-\frac{1}{2}\alpha_2(\alpha_2-4\tau)}-e^{-\frac{1}{2}\alpha_1(\alpha_1+4\tau)}-e^{-\frac{1}{2}\alpha_2(\alpha_2+4\tau)}\Big)\\
&\hspace{108.5pt}-\frac{1}{4}\Big(e^{-\frac{1}{2}(\alpha_1-\alpha_2)(\alpha_1-\alpha_2-4\tau)}-e^{-\frac{1}{2}(\alpha_1+\alpha_2)(\alpha_1+\alpha_2-4\tau)}+e^{-\frac{1}{2}(\alpha_1-\alpha_2)(\alpha_1-\alpha_2+4\tau)}-e^{-\frac{1}{2}(\alpha_1+\alpha_2)(\alpha_1+\alpha_2+4\tau)}\Big)\\
&\hspace{108.5pt}+\sum_{n=1}^\infty\frac{(-1)^n}{1-4n^2}\Big[\frac{4}{\pi^2}\Big(e^{-\frac{1}{2}\alpha_1n(\alpha_1n-4\tau)}+e^{-\frac{1}{2}\alpha_1n(\alpha_1n+4\tau)}+e^{-\frac{1}{2}\alpha_2n(\alpha_2n-4\tau)}+e^{-\frac{1}{2}\alpha_2n(\alpha_2n+4\tau)}\Big)\\
&\hspace{173.5pt}-\frac{1}{\pi}\Big(e^{-\frac{1}{2}(\alpha_2-\alpha_1n)(\alpha_2-\alpha_1n-4\tau)}+e^{-\frac{1}{2}(\alpha_2+\alpha_1n)(\alpha_2+\alpha_1n-4\tau)}\\
&\hspace{193.5pt}-e^{-\frac{1}{2}(\alpha_2-\alpha_1n)(\alpha_2-\alpha_1n+4\tau)}-e^{-\frac{1}{2}(\alpha_2+\alpha_1n)(\alpha_2+\alpha_1n+4\tau)}\\
&\hspace{193.5pt}+e^{-\frac{1}{2}(\alpha_1-\alpha_2n)(\alpha_1-\alpha_2n-4\tau)}+e^{-\frac{1}{2}(\alpha_1+\alpha_2n)(\alpha_1+\alpha_2n-4\tau)}\\
&\hspace{193.5pt}-e^{-\frac{1}{2}(\alpha_1-\alpha_2n)(\alpha_1-\alpha_2n+4\tau)}-e^{-\frac{1}{2}(\alpha_1+\alpha_2n)(\alpha_1+\alpha_2n+4\tau)}\Big)\Big]\\
&\hspace{108.5pt}+\frac{4}{\pi^2}\sum_{n,m=1}^\infty\frac{(-1)^{n+m}}{(1-4n^2)(1-4m^2)}\Big(e^{-\frac{1}{2}(\alpha_2m-\alpha_1n)(\alpha_2m-\alpha_1n-4\tau)}+e^{-\frac{1}{2}(\alpha_2m+\alpha_1n)(\alpha_2m+\alpha_1n-4\tau)}\\
&\hspace{250pt}+e^{-\frac{1}{2}(\alpha_2m-\alpha_1n)(\alpha_2m-\alpha_1n+4\tau)}+e^{-\frac{1}{2}(\alpha_2m+\alpha_1n)(\alpha_2m+\alpha_1n+4\tau)}\Big)\Big\},
\end{split}
\label{zigzag_nonlocal_nonconstant_deco}
\end{equation}
\begin{equation}
\begin{split}
\kappa_{12}(\tau)\big\vert_{K=-1}&=e^{i2\eta\tau}\Lambda_{12}(\tau)\big\vert_{K=1}\\
&=e^{i2\eta\tau}\Big\{\frac{4}{\pi^2}+\frac{1}{2}\Big(e^{-\frac{1}{2}(\alpha_1-\alpha_2)^2}-e^{-\frac{1}{2}(\alpha_1+\alpha_2)^2}\Big)\\
&\hspace{40pt}+\frac{8}{\pi^2}\Big[\sum_{n=1}^\infty\frac{(-1)^n}{1-4n^2}\Big(e^{-2(\alpha_1n)^2}+e^{-2(\alpha_2n)^2}\Big)\\
&\hspace{70pt}+\sum_{n,m=1}^\infty\frac{(-1)^{n+m}}{(1-4n^2)(1-4m^2)}\Big(e^{-2(\alpha_2m-\alpha_1n)^2}+e^{-2(\alpha_2m+\alpha_1n)^2}\Big)\Big]\Big\}.
\end{split}
\label{zigzag_nonlocal_constant_deco}
\end{equation}


\subsection*{\label{sec:appendix_b}Appendix B: Solving the decoherence functions using the parabola phase distribution}

\setcounter{equation}{0}
\renewcommand{\theequation}{B\arabic{equation}}

Here, we make use of the identity
\begin{equation}
\int dxe^{i\frac{1}{2}Ax^2+iBx}=\sqrt{\frac{i2\pi}{A}}e^{-i\frac{B^2}{2A}}.
\label{parabola_phase_identity}
\end{equation}
Thus, the local decoherence functions become
\begin{equation}
\begin{split}
\kappa_j(\tau)&=e^{i\eta\tau}\int dx_j\frac{1}{\sqrt{2\pi}}e^{-\frac{1}{2}x_j^2}e^{i\beta_jx_j^2}e^{i\tau x_j}\\
&=e^{i\eta\tau}\int dx_j\frac{1}{\sqrt{2\pi}}e^{i\frac{1}{2}(2\beta_j+i)x_j^2+i\tau x_j}\\
&=\sqrt{\frac{i}{2\beta_j+i}}e^{i\big(\eta\tau-\frac{1}{2(2\beta_j+i)}\tau^2\big)}.
\end{split}
\label{parabola_kappa_j}
\end{equation}
As for the absolute values, we obtain the Gaussians
\begin{equation}
|\kappa_j(\tau)|=(4\beta_j^2+1)^{-\frac{1}{4}}e^{-\frac{1}{2}\frac{1}{4\beta_j^2+1}\tau^2}.
\label{abs_parabola_kappa_j}
\end{equation}
The non-local decoherence functions are obtained in a similar fashion. Eq.~\eqref{parabola_phase_identity} is used twice in a row. Here, we keep the correlation coefficient $K$ free. The resulting decoherence functions are
\begin{equation}
\kappa_{12}(\tau)=\sqrt{\frac{1}{1-4\beta_1\beta_2(1-K^2)+i2(\beta_1+\beta_2)}}e^{-\frac{1+K+i(\beta_1+\beta_2)(1-K^2)}{1-4\beta_1\beta_2(1-K^2)+i2(\beta_1+\beta_2)}\tau^2+i2\eta\tau},
\label{parabola_kappa_12}
\end{equation}
\begin{equation}
\Lambda_{12}(\tau)=\sqrt{\frac{1}{1+4\beta_1\beta_2(1-K^2)+i2(\beta_1-\beta_2)}}e^{-\frac{1-K+i(\beta_1-\beta_2)(1-K^2)}{1+4\beta_1\beta_2(1-K^2)+i2(\beta_1-\beta_2)}\tau^2}.
\label{parabola_lambda_12}
\end{equation}
\\
The non-local oscillations emerge, when we employ the frequency distribution $\frac{1}{2}[G(1)+G(-1)]=:\mathbf{X}$. The magnitudes of Eqs. \eqref{parabola_kappa_12} and \eqref{parabola_lambda_12} then become
\begin{equation}
\begin{split}
|\kappa_{12}(\tau)|_\mathbf{X}&=\frac{1}{2}\text{Abs}\Big(\kappa_{12}(\tau)\big\vert_{K=1}+\kappa_{12}(\tau)\big\vert_{K=-1}\Big)\\
&=\frac{1}{2}[1+4(\beta_1+\beta_2)^2]^{-\frac{1}{4}}\sqrt{1+2e^{-\frac{2}{1+4(\beta_1+\beta_2)^2}\tau^2}\cos\Big(\frac{4(\beta_1+\beta_2)}{1+4(\beta_1+\beta_2)^2}\tau^2\Big)+e^{-\frac{4}{1+4(\beta_1+\beta_2)^2}\tau^2}},
\end{split}
\label{abs_parabola_kappa_12}
\end{equation}

\begin{equation}
\begin{split}
|\Lambda_{12}(\tau)|_\mathbf{X}&=\frac{1}{2}\text{Abs}\Big(\Lambda_{12}(\tau)\big\vert_{K=1}+\Lambda_{12}(\tau)\big\vert_{K=-1}\Big)\\
&=\frac{1}{2}[1+4(\beta_1-\beta_2)^2]^{-\frac{1}{4}}\sqrt{1+2e^{-\frac{2}{1+4(\beta_1-\beta_2)^2}\tau^2}\cos\Big(\frac{4(\beta_1-\beta_2)}{1+4(\beta_1-\beta_2)^2}\tau^2\Big)+e^{-\frac{4}{1+4(\beta_1-\beta_2)^2}\tau^2}}.
\end{split}
\label{abs_parabola_lambda_12}
\end{equation}


\twocolumngrid


\begin{thebibliography}{99}

\bibitem{petru}
H.-P. Breuer and F. Petruccione, \emph{The Theory of Open Quantum Systems} (Oxford University Press, Oxford, 2002).

\bibitem{suter}
D. Suter and G. A. \'Alvarez, Rev. Mod. Phys.  \textbf{88}, 041001 (2016).

\bibitem{RevRivas}
\'A. Rivas, S. F. Huelga, and  M. B. Plenio,
Rep. Prog. Phys. {\bf 77},  094001 (2014).

\bibitem{RevRMP}
H.-P. Breuer, E.-M. Laine, J. Piilo, and B. Vacchini,
Rev. Mod. Phys. {\bf 88}, 021002 (2016).

\bibitem{devega}
I. de Vega and D. Alonso,
Rev. Mod. Phys. {\bf 89}, 015001 (2017).

\bibitem{LiRev}
L. Li, M. J. W. Hall, and H. M.  Wiseman,
Phys. Rep. {\bf 759} 1 (2018).

\bibitem{p1}
C.-F. Li, G.-C. Guo, and J. Piilo,
EPL (Europhys. Lett.) \textbf{127}, 50001 (2019).

\bibitem{p2}
C.-F. Li, G.-C. Guo, and J. Piilo,
EPL (Europhys. Lett.) \textbf{128}, 30001 (2019).

\bibitem{telep}
E.-M. Laine, H.-P. Breuer, and J. Piilo, 
Sci. Rep.  \textbf{4}, 4620 (2014).

\bibitem{SDC}
B.-H. Liu {\it{et al.}}, EPL (Europhys. Lett.) \textbf{114}, 10005 (2015).

\bibitem{josza}
Y. Dong {\it{et al.}}, npj Quantum Inf. \textbf{4}, 3 (2018).

\bibitem{from_M_to_NM}
B.-H. Liu, L. Li, Y.-F. Huang, C.-F. Li, G.-C. Guo, E.-M. Laine, H.-P. Breuer, and J. Piilo, Nat. Phys. \textbf{7}, 931 (2011).

\bibitem{dephasing_control}
Z.-D. Liu, H. Lyyra, Y.-N. Sun, B.-H. Liu, C.-F. Li, G.-C. Guo, S. Maniscalco, and J. Piilo, Nat. Commun. \textbf{9}, 3453 (2018).

\bibitem{tempo}
A. Strathearn, P. Kirton, D. Kilda, J. Keeling, and B. W. Lovett, Nat. Commun. \textbf{9}, 3322 (2018).

\bibitem{NM_to_M_1}
A. Imamoglu, Phys. Rev. A \textbf{50}, 3650 (1994).

\bibitem{NM_to_M_2}
B. J. Dalton, S. M. Barnett, and B. M. Garraway, Phys. Rev. A \textbf{64},
053813 (2001).

\bibitem{NM_to_M_3}
R. Martinazzo, B. Vacchini, K. H. Hughes, and I. Burghardt, J. Chem. Phys. \textbf{134}, 011101 (2011).

\bibitem{Chiuri}
A. Chiuri {\it{et al.}}, Sci. Rep. \textbf{2}, 968 (2012).

\bibitem{Cialdi}
S. Cialdi {\it{et al.}}, Appl. Phys. Lett. \textbf{10}, 081107 (2017).

\bibitem{fff}
F. F. Fanchini {\it{et al.}}, Phys. Rev. Lett. \textbf{112}, 210402 (2014).

\bibitem{bnk}
N.K. Bernardes {\it{et al.}}, Sci. Rep. \textbf{5}, 17520 (2015).

\bibitem{spectra}
S. Yu {\it{et al.}}, Phys. Rev. Lett. \textbf{120}, 060406 (2018).

\bibitem{collision}
\'A. Cuevas {\it{et al.}}, Sci. Rep. \textbf{9}, 3205 (2019).

\bibitem{nonlocal_memory}
E.-M. Laine, H.-P. Breuer, J. Piilo, C.-F. Li, and G.-C. Guo, Phys. Rev. Lett. \textbf{108}, 210402 (2012); {\it{ibid.}}  \textbf{111}, 229901 (2013).

\bibitem{photonic_real}
B.-H. Liu, D.-Y. Cao, Y.-F. Huang, C.-F. Li, G.-C. Guo, E.-M. Laine, H.-P. Breuer, and J. Piilo, Sci. Rep. \textbf{3}, 1781 (2013).

\bibitem{K_approx_one}
F. Laudenbach, H. Hübel, M. Hentschel, P. Walther, and A. Poppe, Opt. Express \textbf{24}, 2712 (2016).

\bibitem{x_state}
T. Yu and J. Eberly, Quantum Inf. Comput. \textbf{7}, 459 (2007).

\bibitem{teiko}
T. Heinosaari and M. Ziman, \emph{The Mathematical Language of Quantum Theory} (Cambridge University Press, Cambridge, 2012).

\bibitem{dfs1}
D. Lidar, Adv. Chem. Phys. \textbf{154}, 295 (2014).

\bibitem{dfs2}
J. Altepeter, P. Hadley, S. Wendelken, A. Berglund, and P. Kwiat, Phys. Rev. Lett. \textbf{92}, 147901 (2004).

\bibitem{quantum_correlation}
Z. Xi, Y. Li, and H. Fan, Sci. Rep. \textbf{5}, 10922 (2015).

\bibitem{entropy}
S. Luo, S. Fu, and H. Song, Phys. Rev. A \textbf{86}, 044101 (2012).

\bibitem{chuang}
M. Nielsen and I. Chuang, \emph{Quantum Computation and Quantum Information} (Cambridge University Press, Cambridge, 2000).

\bibitem{fidelity}
R. Jozsa, J. Mod. Opt. \textbf{41}, 2315 (1994).

\end{thebibliography}
\end{document}